\documentclass[]{aa}
\usepackage{graphicx}	% Including figure files

\usepackage{amsmath}	% Advanced maths commands
\usepackage{amssymb}	% Extra maths symbols
\usepackage{multicol}   % Multi-column entries in tables
\usepackage{color}
\usepackage{xspace}
\usepackage{subfigure}
\usepackage{CJKutf8}
\usepackage{txfonts}
\usepackage{natbib,twoopt}
\usepackage[colorlinks=true,
            linkcolor=blue,
            urlcolor=blue,
            citecolor=blue,
            anchorcolor=blue]{hyperref}
\makeatletter
\renewcommand*\aa@pageof{, page \thepage{} of \pageref*{LastPage}}
\makeatother

\definecolor{mygray}{gray}{0.6}
\definecolor{TsinghuaPurple}{cmyk}{0.58,0.90,0,0}
\definecolor{magenta}{rgb}{0.858, 0.188, 0.478}

\definecolor{mygray}{gray}{0.6}
\definecolor{emerald}{RGB}{0,155,155}
\newcommand{\hjc}[1]{\textcolor{emerald}{[\textit{\small #1}]}}

\newcommand{\hjrem}[1]{\textcolor{mygray}{\sout{#1}}}
\newcommand{\xxx}[1]{\textcolor{blue}{\textbf{xxx}\xspace}}
\newcommand{\fg}[1]{Fig.~\ref{fig:#1}}
\newcommand{\eq}[1]{Eq.~(\ref{eq:#1})\xspace}
\newcommand{\Eq}[1]{Equation~(\ref{eq:#1})\xspace}%beginning of the sentence

\newcommand{\tb}[1]{Table~\ref{tab:#1}\xspace}
\newcommand{\Tb}[1]{Table~\ref{tab:#1}\xspace}%beginning of the sentence

\newcommand{\App}[1]{Appendix~\ref{app:#1}\xspace}

\begin{document}

    \title{Grain-size measurements in protoplanetary disks indicate fragile pebbles and low turbulence}
    %\subtitle{A Coincidence or a Correlation?}
    \titlerunning{fragile pebbles in low-turbulence disks}

    \author{Haochang Jiang (\begin{CJK*}{UTF8}{gbsn}蒋昊昌\end{CJK*})
    \inst{1}\fnmsep\inst{2} %\fnmsep\inst{\star}
    \and
    Enrique Mac\'{i}as\inst{1}
    \and
    {Osmar M. Guerra-Alvarado\inst{3}}
    \and
    {Carlos Carrasco-Gonz{\'a}lez\inst{4}}
    }
    \authorrunning{H. Jiang, E. Mac\'{i}as, O.~M. Guerra-Alvarado, C. Carrasco-Gonz{\'a}lez}

    \institute{
    European Southern Observatory,
    Karl-Schwarzschild-Str 2, 85748 Garching, Germany\\
    \email{\href{mailto:Haochang.Jiang@eso.org}{hjiang@eso.org}}
    \and
    Department of Astronomy, Tsinghua University,
    30 Shuangqing Rd, Haidian DS 100084, Beijing, PR China
    \and
    {Leiden Observatory, Leiden University, PO Box 9513, 2300 RA Leiden, The Netherlands}
    \and
    {Instituto de Radioastronomía y Astrofísica (IRyA), Universidad Nacional Autónoma de México (UNAM), Mexico}
    }

   \date{Received September 15, 1996; accepted March 23, 1997}

% \abstract{}{}{}{}{} 
% 5 {} token are mandatory
 
  \abstract
   {Constraining the turbulence level and pebble size in protoplanetary disks is an essential initial step in understanding the aerodynamic properties of pebbles, which are crucial for planet formation. Recent laboratory experiments have revealed that destructive collisions of icy dust particles {may} occur at much lower velocities than previously believed. These low fragmentation velocities push down the maximum grain size in collisional growth models.}
   {Motivated by the smooth radial distribution of pebble sizes inferred from ALMA/VLA multi-wavelength continuum analysis, we propose a concise model to explain this feature and aim to constrain the turbulence level at the midplane of protoplanetary disks.}
   {Our approach is built on the assumption that the fragmentation threshold is the primary barrier limiting pebble growth within pressure maxima. Consequently, the grain size at the ring location can provide direct insights into the turbulent velocity governing pebble collisions and, by extension, the turbulence level at the disk midplane. We validate this method using the \texttt{Dustpy} code, which simulates dust transport and coagulation.}
   {We apply our method to 7 disks, TW~Hya, IM~Lup, GM~Aur, AS~209, HL~Tau, HD~163296, and MWC~480, for which grain sizes have been measured from multi-wavelength continuum analysis. A common feature emerges from our analysis, with an overall low turbulence coefficient of $\alpha\sim10^{-4}$ observed in five out of seven disks when taking fragmentation velocity $v_{\rm frag} = 1{\rm \,m\,s}^{-1}$. A higher fragmentation velocity would imply a turbulence coefficient significantly larger than the current observational constraints. IM~Lup stands out with a relatively higher coefficient of $10^{-3}$. Notably, HL~Tau exhibits an increasing trend in $\alpha$ with distance, {which supports enhanced turbulence at its outer disk region, possibly associated with the infalling streamer onto HL~Tau. Alternatively, if the turbulence was low, it might indicate} that grain sizes have not reached the growth barrier.}% due to the young Class~I nature of HL~Tau.}
   {We conclude that the current (sub)mm pebble size constrained in disks implies low levels of turbulence, as well as fragile pebbles consistent with recent laboratory measurements.}


   \maketitle
%
%-------------------------------------------------------------------

\section{Introduction}\label{sec:introduction}
(Sub)mm size dust particles, commonly referred to as pebbles, are the basic building blocks of planets. To understand the process of planet formation, it is essential to comprehend pebble properties, their coagulation growth, and their aerodynamics.

In protoplanetary disks, the rotation of gas deviates from Keplerian velocities in the disk due to the pressure support, the pebbles are therefore dragged by the gas, changing their angular momentum, which leads to a radial drift \citep{Weidenschilling1977a}. As the gas is viscous in the protoplanetary disk, the pebbles in the disk also diffuse along with gas turbulent mixing \citep[e.g.,][]{OrmelCuzzi2007,YoudinLithwick2007}. These effects all depend on the pebble size, which evolves depending on the balance between sticking collisions, fragmentation \citep[e.g.,][]{BrauerEtal2008,BirnstielEtal2010}, bouncing \citep[e.g.,]{ZsomEtal2010,WindmarkEtal2012a} and pebble radial drift \citep[e.g.,][]{BirnstielEtal2012}.

Thanks to the development of advanced radio facilities like the Atacama Large Millimeter/submillimeter Array (ALMA) and the Karl G. Jansky Very Large Array (VLA), we are now in an era where pebble properties can be characterized by their thermal emission continuum on an astronomical unit (au) scale. 
Multi-wavelength observations can be used to infer the maximum size and size distribution of pebbles \citep[e.g., ][]{Carrasco-GonzalezEtal2019,MaciasEtal2019,MaciasEtal2021,SierraEtal2021,GuidiEtal2022}. This method capitalizes on the connection between the frequency dependence of dust opacity and the particle size distribution. Particularly, in the optically thin regime, such a relationship is relatively straightforward within the Rayleigh-Jeans regime: the spectral index of (sub)millimeter dust emission aligns with 2 + $\beta$, where $\beta$ signifies the spectral index of dust opacity \citep{BeckwithEtal1990}. Smaller values of $\beta$ indicate the presence of larger dust particles \citep{MiyakeNakagawa1993,D'AlessioEtal2001}. In addition, ALMA is also revolutionizing our view of the gas disk, mostly coming from the high resolution and high sensitivity observation on CO isotopologues \citep[e.g.,][and reference therein]{ZhangEtal2019,BoothIlee2020,CalahanEtal2021a,ZhangEtal2021k,MiotelloEtal2023}. Based on the pebble sizes $a_{\rm p}$ and surface densities $\Sigma_{\rm g}$ inferred from ALMA observations, the so-called Stokes parameter in the Epstein regime can be expressed as
\begin{equation}\label{eq:St}
    {\rm St} = \frac{\pi}{2} \frac{\rho_\bullet a_{\rm p}}{\Sigma_{\rm g}}
\end{equation}
where $\rho_\bullet$ is the internal density of pebbles, quantifying the aerodynamic coupling between gas and dust. 

At the same time, laboratory experiments provide us with different insights into pebble properties, particularly regarding pebble collision \citep[see the recent review of ][]{WurmTeiser2021}. Historically, collision experiments involving small micrometer-sized water ice grains have demonstrated sticking behavior even at velocities of $10$\,m\,s$^{-1}$, whereas silicates typically exhibit sticking behavior at $1$\,m\,s$^{-1}$ \citep[e.g.,][]{PoppeEtal2000,BlumWurm2008}. Additionally, simulations conducted by \citet{DominikTielens1997} have indicated that ice aggregates possess greater stability compared to silicate aggregates. However, recently, it has been proposed that water ice exhibits reduced stickiness at lower temperatures \citep{GundlachEtal2018}. In laboratory experiments, the surface energy of water ice remains relatively high between the freezing point of water (273~K) and 200~K \citep{GundlachEtal2011,AumatellWurm2014}. However, the surface energy drops by two orders of magnitude below 200~K, which is closer to the range of temperatures found in the outer region of protoplanetary disks \citep{MusiolikWurm2019}. It also remains unclear how the condensation of other molecules on the pebble surface might change the collision properties \citep{HommaEtal2019,BischoffEtal2020}. While CO$_2$, one of the prevalent ices in protoplanetary disks, has been investigated in collision experiments, it has been demonstrated to be as or less adhesive than water ice or silicates \citep{MusiolikEtal2016a}. 
%Therefore, fragile pebbles seem expected throughout the outer disk region.

To understand dust evolution and planet formation, another crucial disk property is the magnitude of the gas turbulence, and therefore the dust diffusivity. This parameter partly sets the relative velocity among pebbles \citep[][]{OrmelCuzzi2007}, which directly decides their collision outcomes \citep{GuettlerEtal2010,ZsomEtal2010,ZsomEtal2011}.
There have been various attempts to measure the gas viscosity and/or dust diffusivity in protoplanetary disks using different observables: the disk vertical extent in ALMA continuum as a proxy of dust settling \citep[e.g.,][]{PinteEtal2016,VillenaveEtal2020}, the balance between dust ring width and pressure gradient support \citep[e.g.,][]{DullemondEtal2018,RosottiEtal2020}, non-thermal broadening of emission lines \citep[e.g.,][]{TeagueEtal2018,FlahertyEtal2020}, and also from the disk evolution constraints set by demographic studies (see the recent reviews by \citealp{MiotelloEtal2023} and \citealp{Rosotti2023}). Multiple lines of evidence suggest that discs are in fact not as turbulent as the previously assumed value of $\alpha = 10^{-2}$ \citep{Rosotti2023}.

In this paper, we propose that the pebble sizes inferred from multi-wavelength observations can be used to put constraints on the diffusivity and fragmentation velocity of dust particles. First, we briefly discuss the benchmark test model we run with the newly developed \texttt{Dustpy} code \citep{StammlerBirnstiel2022} and explain the rationale. We then apply our proposed new method to seven previously analyzed protoplanetary disks, finding values of the dust diffusivity that are consistent with previous estimates of disc turbulence, as long as the fragmentation velocity is low even when coated by water ice.

\section{Model}\label{sec:model}
To examine the impact of diffusivity and fragmentation velocity on grain growth within the pebble ring, we employ the open-source code \texttt{DustPy}\footnote{We refer to the \href{https://stammler.github.io/dustpy/index.html}{\texttt{DustPy} documentation} for details of the code.} \citep{StammlerBirnstiel2022} to model dust coagulation and transport in protoplanetary disks. {This code concurrently addresses the Smoluchowski coagulation equation \citep{Smoluchowski1916} along with dust particle transport within the disk. Dust particles undergo collisions and are transported by various mechanisms, including thermal Brownian motion, vertical stirring and settling, turbulent mixing, as well as azimuthal and radial drift. The code considers all these velocities when calculating relative velocities prior to collision.} Our simulations incorporate various disk parameters, resulting in limitations on dust growth either due to drifting or fragmentation at the output snapshots.

We initialize the gas density following the similarity solution of \citet{Lynden-BellPringle1974}
\begin{equation}\label{eq:Sig}
\begin{aligned}
    \Sigma_{\rm g} &= \Sigma_{\rm g,0}\left(\frac{r}{r_c}\right)^{-\gamma}
    \exp\left[-\left(\frac{r}{r_c}\right)^{2-\gamma}\right]
    \\
\end{aligned}
\end{equation}
with $r$ the distance from the star, $\gamma = 1$, and a characteristic radius of $r_c = 100$~au throughout this work. {The gas surface densities evolve with the master equation of viscous accretion disk \citep{Lynden-BellPringle1974}:
\begin{equation}
    \frac{\partial \Sigma_{\rm g}}{\partial t} 
    = -\frac{1}{r}\frac{\partial}{\partial r}(\Sigma_{\rm g}r v_{\rm g})
    \equiv \frac{3}{r}\frac{\partial}{\partial r}\left[\frac{1}{r\Omega_K}\frac{\partial}{\partial r}(\alpha_v\Sigma_{\rm g}r^2c_s^2)\right]
\end{equation}
where $\Omega_K \equiv v_K/r = \sqrt{GM_\star/r^3}$ is the Keplerian orbital frequency around the star with a mass of $M_\star$, $\alpha_v$ is the turbulent viscosity coefficient defined by the \citet{ShakuraSunyaev1973} prescription, and $c_s$ is the sound speed. 
}
The disk is assumed to be locally isothermal, with the disk temperature following:
\begin{equation}\label{eq:Td}
    T_{\rm d} = T_0\times\left(\frac{r}{r_c}\right)^{-q}.
\end{equation} 
We take $T_0 = 20\,K$, $q=0.45$, with which, the aspect ratio of gas disk follow
\begin{equation}\label{eq:cs}
    h_{\rm g} \equiv \frac{H_\mathrm{g}}{r} = \frac{c_s}{\Omega_K r} = h_0\times\left(\frac{r}{r_c}\right)^{0.5-q/2}
\end{equation}
where $h_0 = 0.075$ is the aspect ratio at $r_c$.

{The dust surface evolution is solved by the advection-diffusion equation \citep{BirnstielEtal2010}
\begin{equation}
    \frac{\partial \Sigma_{\rm d}}{\partial t} +
    \frac{1}{r}\frac{\partial}{\partial r}(\Sigma_{\rm d}r v_{\rm d}) -
    \frac{1}{r}\frac{\partial}{\partial r}\left[r\Sigma_{\rm g}D_{\rm d}\frac{\partial}{\partial r}(\frac{\Sigma_{\rm d}}{\Sigma_{\rm g}})\right] = 0
\end{equation}
with the dust diffusivity $D_{\rm d} = \alpha_v c_s h / (1+{\rm St}^2)$ \citep{YoudinLithwick2007} and the dust radial velocity \citep{NakagawaEtal1986}
\begin{equation}
    v_{\rm d} = \frac{v_{\rm g}}{1+{\rm St}^2} - \frac{2{\rm St}}{1+{\rm St}^2}\eta v_K
\end{equation}
where the dimensionless pressure gradient coefficiency reads
\begin{equation}\label{eq:eta_r}
    \eta = -\frac{1}{2}\frac{c_s^2}{v_K^2} \frac{d {\rm ln}P}{d {\rm ln}r}.
\end{equation}
in which $P$ is the midplane gas pressure.}

We also test the case where pebbles are trapped inside pressure maximum at distant radii. Following \cite{DullemondEtal2018}, we modify the gas viscosity
\begin{equation}
    \alpha_{v,r}(r) = \frac{\alpha_v}{F(r)}
\end{equation}
where
\begin{equation}\label{eq:F_gap}
    F(r) = \exp{\left[-A \exp{\left(-\frac{(r-r_0)^2}{2 w^2}\right)}\right]}
\end{equation}
We insert two viscosity gaps at $r_0 = 40$~au and 90~au separately with gap width $w = 4$~au and 9~au. We take the gap depth amplitude $A=1$ as the default value but try $A= 0.5$ and 2 for parameter studies in \App{add_model}. 

The grain growth is mainly limited by two barriers \citep{BirnstielEtal2010}. The fragmentation barrier of pebble growth when the relative velocity is dominated by turbulent velocity \citep{OrmelCuzzi2007} reads as:
\begin{equation}\label{eq:St_frag}
    {\rm St}_{\rm frag} = \frac{1}{3}\frac{v_{\rm frag}^2}{\delta_{\rm t}c_s^2}
\end{equation}
{where $\delta_{\rm t}$ is the dust turbulent velocity coefficiency, whose value is equal to $\alpha_v$ through the main text.}
And the drifting barrier, balancing drifting timescale and coagulation timescale, is \citep[e.g.,][]{BirnstielEtal2012}:
\begin{equation}\label{eq:St_drift}
    {\rm St}_{\rm drift} = \frac{\epsilon}{2\eta}
\end{equation}
where the barrier is proportional to the local dust-to-gas surface density ratio
$\epsilon = \Sigma_{\rm d}/\Sigma_{\rm g}$,
and inversely proportional to the dimensionless pressure gradient coefficiency $\eta$ (\eq{eta_r}).

\begin{table}
\caption{Summary of disk parameters used in this work and references}
\label{tab:disk_parameters} 
\centering
\small
\begin{tabular}{c|ccc}
\hline\hline
Source & $M_\star$\tablefootmark{a} [$M_\odot$] & dust & gas \\
\hline
TW~Hya    &   0.6   &  (1)  &  (4) \\
GM~Aur    &   1.1   &  (2)  &  (5) \\
IM~Lup    &   1.1   &  (2)  &  (5) \\
AS~209    &   1.2   &  (2)  &  (5) \\
HL~Tau    &   1.7   &  (3)  &  (6) \\
HD~163296 &   2.0   &  (2)  &  (5) \\
MWC~480   &   2.1   &  (2)  &  (5) \\
\hline
\end{tabular}
% 60, 158, 156, 121, 147, 101, 156
\tablefoot{References: (1) \citet{MaciasEtal2021} (2) \citet{SierraEtal2021} (3) Guerra-Alvarado~et~al.~(in~prep.) (4) \citet{CalahanEtal2021a} (5) \citet{ZhangEtal2021k} (6) \citet{BoothIlee2020}\\
\tablefoottext{a}{All stellar masses are dynamically determined by disk Keplerian rotation, TW~Hya \citep{TeagueEtal2022}, HL~Tau \citep{PinteEtal2016}, MAPS sources \citep{OebergEtal2021,TeagueEtal2021}}
}
\end{table}

\subsection{Benchmark simulations}

\begin{figure*}[tbp]
\centering  
\subfiguretopcaptrue
\subfigure{
\includegraphics[height=0.383\textwidth]{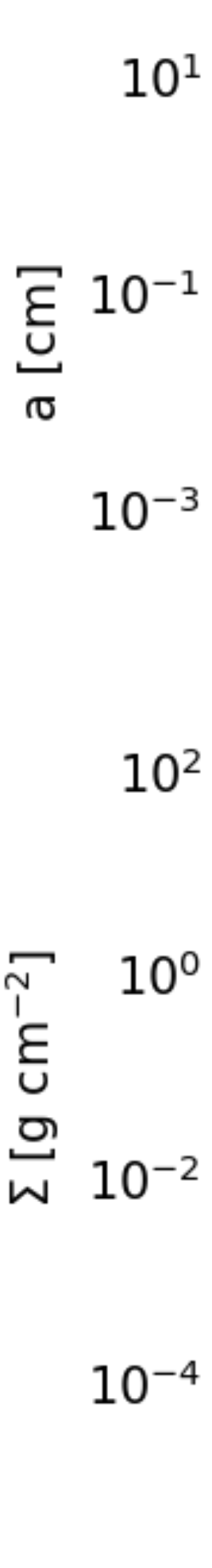}}\hspace{-8pt}
\setcounter{subfigure}{0}
\hfill
\subfigure[$v_{\rm frag} = 10$\,m\,s$^{-1}$]{\label{fig:a4f10_2r_A10}
\includegraphics[height=0.383\textwidth]{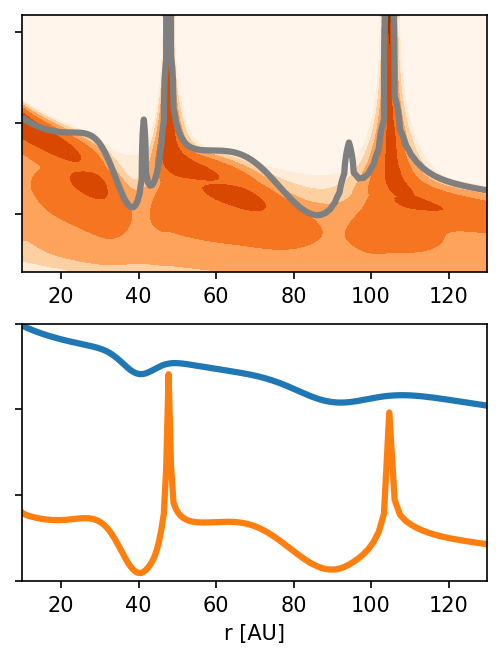}}\hfill\hspace{-10pt}
\subfigure[$v_{\rm frag} = 3$\,m\,s$^{-1}$]{\label{fig:a4f3_2r_A10}
\includegraphics[height=0.383\textwidth]{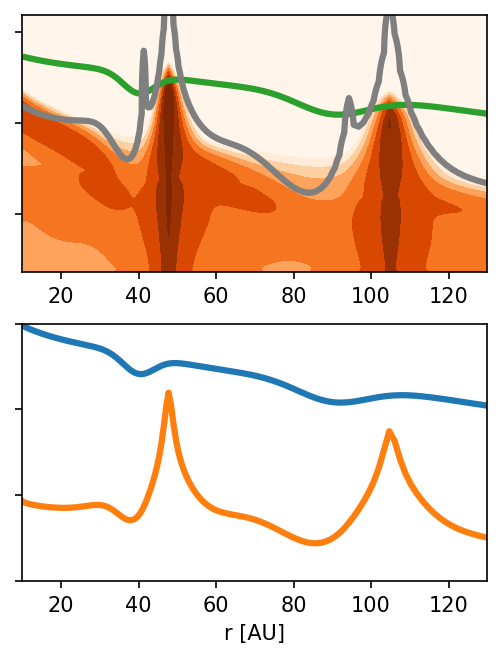}}\hfill\hspace{-10pt}
\subfigure[$v_{\rm frag} = 1$\,m\,s$^{-1}$]{\label{fig:a4f1_2r_A10_1}
\includegraphics[height=0.383\textwidth]{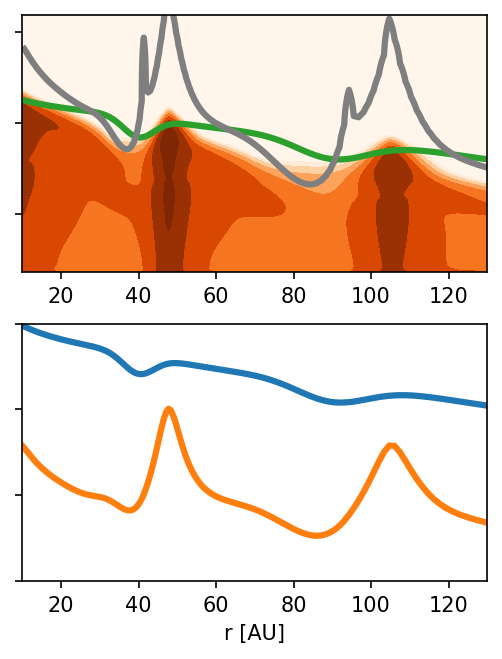}}\hfill\hspace{-10pt}
\subfigure{
\includegraphics[height=0.383\textwidth]{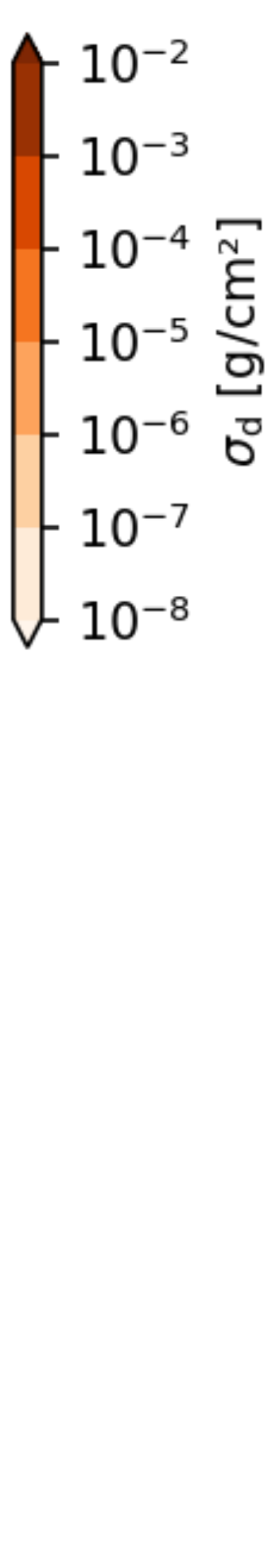}}\hfill
\hfill
\subfiguretopcapfalse
\vspace{-.5 cm}
\caption{
    Simulation outputs at $t=1$~Myr for runs with different fragmentation velocities $v_{\rm frag} = 10, 3, 1\rm\,m\,s^{-1}$ (from left to right respectively). 
    The dust diffusivities are set to be $10^{-4}$ for all three runs. In the upper panels, the gray lines mark the drifting barrier, while the green lines mark the fragmentation barrier in each simulation. 
    The orange contours represent the dust densities at each grain size. 
    The lower panels display the radial profiles of gas (blue) and dust (orange) total column densities. 
    In the case of high fragmentation velocities, particle size increases significantly at the ring location. Yet, for fragile dust ($v_{\rm frag} = 1\rm m\,s^{-1}$), dust growth is limited by fragmentation throughout the entire disk, resulting in a smooth maximum grain size distribution.
    }
\label{fig:dp_v_frag}
\end{figure*}

As a main result, we test the case in which the disk has low viscosity $\alpha_v$ and dust diffusivities $\delta_{\rm t}$. 

In the traditional picture where the icy pebbles are sticky, the fragmentation velocity exterior to the water snowline is expected to be $10$\,m\,s$^{-1}$. In such a case, the grain size is overall controlled by the radial drift of the pebbles. As shown in the \fg{a4f10_2r_A10}, when the fragmentation velocity is relatively high while the dust turbulent velocity is low, the regions of the disk with a smooth pressure gradient cannot host pebbles larger than 1\,mm. However, at the pressure maximum location, the pebbles follow the pressure gradient and are trapped at a specific location, where the radial drift stops. As a result, the so-called drifting barrier no longer limits the pebble growth, and thus pebbles can significantly hit and stick, growing to $>10$\,cm size. 

As the new laboratory results suggest, the pebbles can in turn be very fragile in the outer region of the disk, especially when they are exterior to the CO$_2$ snowline. By varying the fragmentation velocities of the pebble, we test how the fragile nature will influence the size distribution of pebbles in the protoplanetary disks.

%Such growth is expected to exhibit in multi-wavelength observation in the (sub)mm to cm wavelength range. As pebbles grow bigger, the spectral index is expected to change. 
In \fg{a4f3_2r_A10}, with the fragmentation velocity $v_{\rm frag}$ reduced from $10$\,m\,s$^{-1}$ to $3$\,m\,s$^{-1}$, the fragmentation barrier goes down by one order of magnitude (the {green} line in the upper panel). In contrast, the drifting barrier does not significantly change as it is independent of the collisional properties of the pebble. Therefore, even though the fragmentation velocity is lower, the particle size still reaches 1\,mm in the disk region where the density/pressure gradient is smooth. At the ring location, where the radial drift is limited, pebbles can now only grow to ${\sim}$cm size inside the ring with the low fragmentation barrier. 

Further decreasing the fragmentation velocity will result in an even lower fragmentation barrier, which is eventually lower than the nominal drifting barrier in the simulation with $v_{\rm frag} = 1$\,m\,s$^{-1}$. In this run, the fragmentation barrier is lower than the drifting barrier in almost the entire disk. As the \fg{a4f1_2r_A10_1} presents, the maximum pebble is overall constant over the entire disk. 

\begin{figure*}
    \centering
    \includegraphics[width=0.99\textwidth]{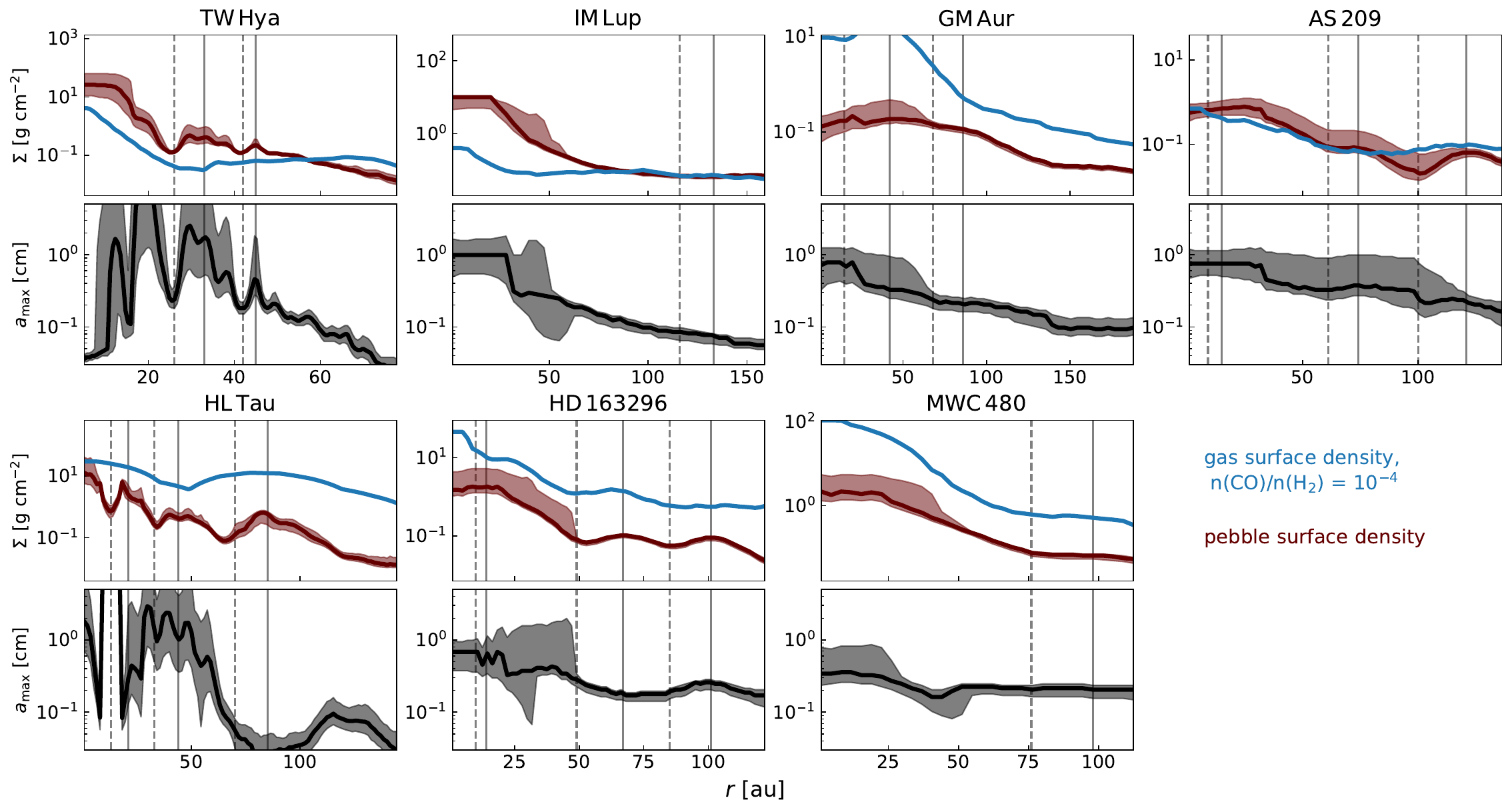}
    \caption{Summary of pebble and gas properties. The pebble surface density (brown) and maximum pebble size (gray) are obtained from multiple-wavelength analysis. The gas surface densities are obtained from modeling based on ALMA observations of CO isotopologues. The blue curves we present here are those by assuming the CO abundance n(CO)/n(H$_2$)$=10^{-4}$, which shall be taken as the lower limit of the total gas densities since significant CO depletion could exist \citep[e.g.,][]{ZhangEtal2021k}. {The peaks of continuum rings and the minima of continuum gaps are indicated by solid and dashed vertical lines separately.}}
    \label{fig:Sig_amax}
\end{figure*}

\begin{figure*}
    \centering
    \includegraphics[width=0.99\textwidth]{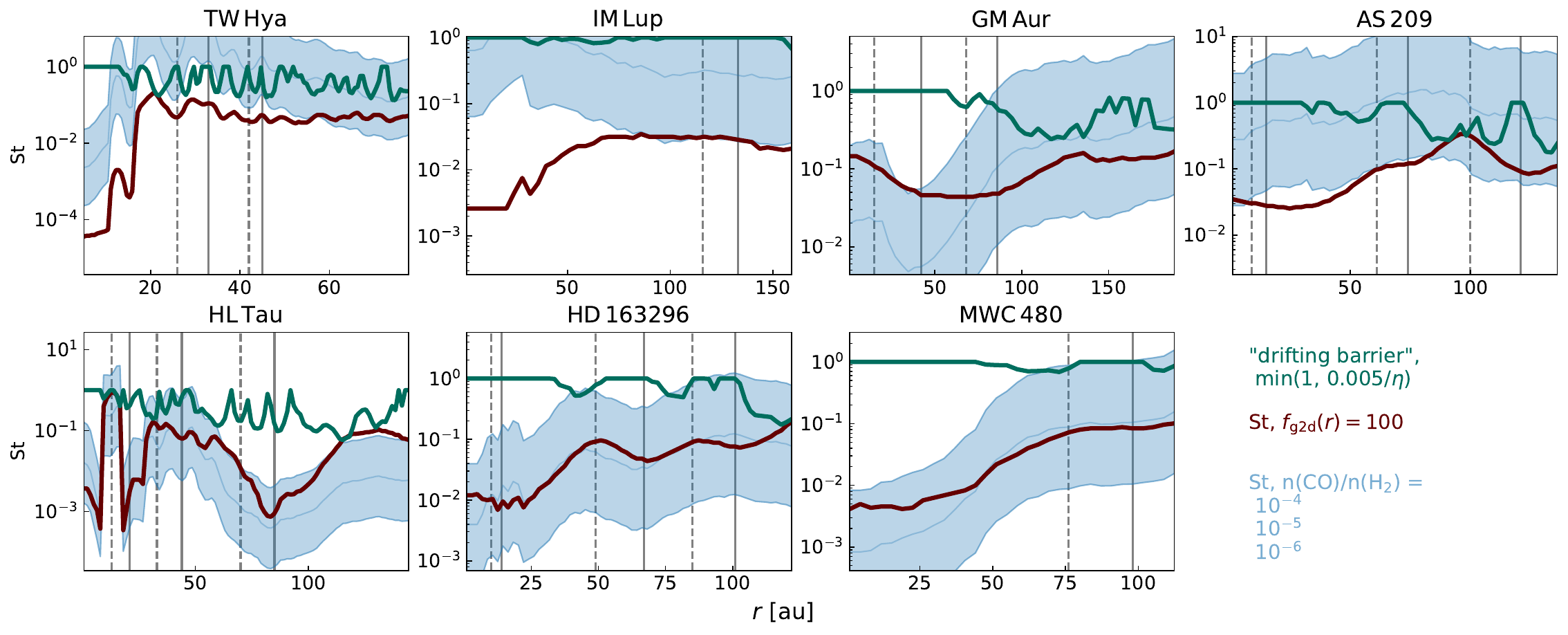}
    \caption{Calculated Stokes number. The blue lines indicate the cases when assuming CO abundance is $10^{-4}$, $10^{-5}$, and $10^{-6}$ from top to bottom. The red lines are values when assuming the gas-to-dust ratio is always 100. The green lines mark the drifting barriers \eq{St_drift} or ${\rm St} = 1$. {Since we assume the gas-to-dust ratio is equal to 100, \Eq{St_drift} can be simply written as ${\rm St}_{\rm drift} = 0.005/\eta$.}}
    \label{fig:St_r}
\end{figure*}

Interestingly, the recent analyses of multi-wavelength observations in several disks have revealed flat radial profiles of the maximum grain size. In particular, \citet{SierraEtal2021} reported an almost constant maximum grain size radial profile across the B49 continuum ring in HD\,163296, without a signal of localized pebble growth. This trend is further confirmed by \citet{GuidiEtal2022}, who suggests that the pebble-size radial profile is smooth over the entire disk, even at the B100 continuum ring. This result is, by and large, consistent with the result we present in low turbulence, low fragmentation velocity simulation (\fg{a4f1_2r_A10_1}). 

Therefore, utilizing the fact that most of the disk habituates a smooth max-grain size profile, we can, in turn, constrain the diffusivity level of the pebble disk by assuming that the maximum pebble size $a_{\rm max}$ is set by the fragmentation barrier, i.e., by assuming ${\rm St} = {\rm St}_{\rm frag}$, we may obtain the $\delta_{\rm t}$ in \eq{St_frag} following:
\begin{equation}\label{eq:alpha_frag}
    \alpha_{\rm frag} \equiv \delta_{\rm t} 
    = \frac{1}{3} \frac{v_{\rm frag}^2}{{\rm St} c_{\rm s}^2},
\end{equation}
a simple formula where both St and $c_s$ could get measured from multi-wavelength continuum analysis.

\section{Results}\label{sec:results}

We apply our analysis on seven disks, where multi-wavelength analysis has been conducted using the same dust opacity \citep[the DSHARP opacity,][]{BirnstielEtal2018}. The sample consists of HL~Tau, TW~Hya, and the five Molecules with ALMA at Planet-forming Scales (MAPS) disks (GM~Aur, IM~Lup, AS~209, HD~163296, and MWC~480). 
The references of the disk information used in this work are listed in \tb{disk_parameters}. 
To summarize, the pebble surface densities $\Sigma_{\rm p}(r)$, the maximum grain size $a_{\rm max}(r)$, and the temperature profiles to estimate $c_s$ are all inferred from multi-wavelength observation analysis. The gas surface densities $\Sigma_{\rm g}(r)$ are based on the intensities of CO isotopologue lines, which assume the abundance of CO the same as the ISM value of n(CO)/n(H$_2$)$=10^{-4}$, in other words, no CO depletion (see more discussion below). In \fg{Sig_amax}, for each disk, pebble densities (red) and gas densities (blue) are plotted in the upper panels. The maximum grain size radial profile $a_{\rm max}(r)$ is presented in the lower panel, where the solid lines are the best fitting results and the shaded regions are $1\sigma$ uncertainties. 
We also mark the peaks of the continuum rings and the minima of continuum gaps by solid and dashed vertical lines separately. 

With these pebble surface densities, we calculate the Stokes number of the maximum-grain-size pebbles using \eq{St} by taking $\rho_\bullet = 1.675$~g cm$^{-3}$ \citep[the DSHARP opacity][]{BirnstielEtal2018}, and the gas surface densities inferred from CO, with n(CO)/n(H$_2$)$=10^{-4},\,10^{-5},\,10^{-6}$. These results are indicated by the blue solid lines in \fg{St_r}, from top down n(CO)/n(H$_2$)$=10^{-4}\,10^{-5},10^{-6}$. The CO in the protoplanetary disk is known to be depleted compared with the ISM CO abundance of $10^{-4}$, typically by one or two orders of magnitudes \citep[e.g.,][]{AnsdellEtal2016,MiotelloEtal2017,KrijtEtal2020,ZhangEtal2021k}. 
Thus, the gas surface densities with n(CO)/n(H$_2$)$=10^{-4}$ used here should be considered as a lower limit, hence the Stokes should be an upper limit.

However, measuring the gas disk mass is a complex task \citep[see Fig.\,10. of ][for a nice illustration on the gas mass of TW~Hya in different methods]{MiotelloEtal2023}. Therefore, in our calculations, we resort to using the Stokes number with reference gas surface densities obtained by assuming the gas surface density is 100 times the pebble surface density (the red lines in \fg{St_r}). This assumption implies that the disk maintains a typical ISM metallicity throughout its extent, a practical and broadly applicable assumption, which is roughly consistent with the ones estimated from the CO gas surface density, assuming a moderate CO depletion \citep{ZhangEtal2021k}. It's important to note that this gas surface density represents an upper limit since the effects of radial drift will likely lead to local gas-to-dust ratios lower than 100. As a result, the Stokes number presented here should be considered lower limits. For clarity, we also include the Stokes number of the drifting barrier (with a cut above St=1 given that gas drag is less important, and these particles would decouple from the turbulence) in green in \fg{St_r}, depending on which value is smaller. It's noteworthy that the drifting barrier is higher than the inferred Stokes number at all radii, suggesting that the pebble size is primarily set by the fragmentation barrier rather than the drifting barrier. {This fact should be specifically robust at the ring location, where the pebbles are trapped and the drifting barriers disappear, as shown in \fg{dp_v_frag}.}

With the Stokes number and fragmentation velocity, we can therefore obtain the $\alpha_{\rm frag}$ by assuming $v_{\rm frag} = 1$\,m\,s$^{-1}$ in \eq{alpha_frag}
\begin{equation}\label{eq:alpha_frag_empir}
\begin{aligned}
    \alpha_{\rm frag}
    &= \frac{2}{3\pi} \frac{f_{\rm g2d}\Sigma_{\rm d}}{\rho_\bullet a_{\rm p}}\frac{v_{\rm frag}^2}{c_{\rm s}^2}\\
    &= 10^{-4} \times 
    \left(\frac{f_{\rm g2d}}{100}\right)
    \left(\frac{\Sigma_{\rm d}}{0.1\,{\rm g\,cm^{-2}}}\right)
    \left(\frac{\rho_\bullet}{1.675\,{\rm g\,cm^{-3}}}\right)^{-1}\\
    &\times
    \left(\frac{a_{\rm p}}{0.3\,\rm cm}\right)^{-1}
    \left(\frac{v_{\rm frag}}{1\,{\rm m\,s^{-1}}}\right)^2
    \left(\frac{c_{\rm s}}{200\,{\rm m\,s^{-1}}}\right)^{-2}
\end{aligned}
\end{equation}
The calculated $\alpha_{\rm frag}$ are presented in \fg{alpha_r}. In \Tb{a_line_frag}, we present a comparison of our findings with two alternative methods based on (sub)millimeter line-broadening and continuum gap contrast. We dashed the $\tau>1$ region with orange lines, where the data in all wavelengths used in the previous analysis becomes optically thick, and therefore precise pebble size could not be measured. Except for GM\,Aur, all the other six sources are optically thick in the inner region, where there might exist degeneracy between larger particles and higher optical depth \citep{RicciEtal2012,Carrasco-GonzalezEtal2019,MaciasEtal2021}. However, we note that the CO$_2$ snowline, which is around 50-70~K, is normally within the inner disk region as well. Therefore, in the optically thin region located outside the CO$_2$ snowline, our assumption of $v_{\rm frag} = 1\rm m\,s^{-1}$ should be robust.

\begin{table*}[htb]
\caption{\label{tab:a_line_frag}
Existing constraints on disk turbulence from sub-mm line broadening $\alpha_{\rm line}$ \citep[][and references therein]{Rosotti2023}, the inferred $\alpha_{\rm frag}$ by assuming pebble size is limited by fragmentation (this work), the $\alpha_{\rm gap}$ constrained from dust settling \citep{DoiKataoka2021,LiuEtal2022y,PizzatiEtal2023}. DM\,Tau and IM\,Lup are the only two sources that show signals of turbulence from line broadening. 
}
\centering
\begin{tabular}{l|cccc}
\hline
Source      & $\alpha_{\rm line}$ & $\alpha_{\rm frag}$ & $\alpha_{\rm frag}/{\rm St}$ & $\alpha_{\rm gap}/{\rm St}$ \\
\hline
TW\,Hya     & $<6\times10^{-3}$   & $3\times10^{-5}$-$2\times10^{-4}$ & $3\times10^{-4}$-$6\times10^{-3}$ &  - \\
IM\,Lup     & 0.03-0.09           & $3\times10^{-4}$-$7\times10^{-4}$ & 0.01-0.03 &  - \\
GM\,Aur     & -                   & $4\times10^{-6}$-$1\times10^{-4}$ & $3\times10^{-5}$-$3\times10^{-3}$ &  - \\
AS\,209     & -                   & $2\times10^{-5}$-$10^{-4}$        & $6\times10^{-5}$-$3\times10^{-3}$ & $\lesssim0.065$ \\
HL\,Tau     & -                   & $7\times10^{-6}$-$3\times10^{-3}$ & $5\times10^{-5}$-$4$ &  - \\
HD\,163296  & $<2.5\times10^{-3}$ & $3\times10^{-5}$-$9\times10^{-5}$ & $1\times10^{-4}$-$2\times10^{-3}$ & $\simeq0.45$ \\
MWC\,480    & $<$0.01             & $5\times10^{-5}$-$9\times10^{-5}$ & $5\times10^{-4}$-$2\times10^{-3}$ &  - \\
%& \multicolumn{4}{l}{\hjc{I guess it might be better to cut off the table from here...}}\\
%V4046\,Sgr  & $<$0.02             & -                                 & -                                 &  - \\
%DM\,Tau     & 0.0625              & -                                 & -                                 &  - \\
%GW\,Lup     & -                   & -                                 & -                                 & $\lesssim0.24$ \\
%DoAr\,25    & -                   & -                                 & -                                 & $\lesssim0.079$ \\
%Elias\,24   & -                   & -                                 & -                                 & $\lesssim0.042$ \\
%MY\,Lup     & -                   & -                                 & -                                 & $\lesssim0.29$ \\
\hline
\end{tabular}
\end{table*}

An evident outcome of our calculations is that, for the majority of the disks, the inferred turbulence level is on the order of $10^{-4}$ in the optically thin region. Furthermore, the observed low turbulence level aligns with constraints derived from alternative methods, as discussed in \citet{Rosotti2023}. Out of the six disks for which line-broadening analysis is available, our study includes four of them, see \Tb{a_line_frag}. 
Notably, turbulence is not clearly detected through line broadening in HD~163296, TW~Hya, and MWC~480, with the exception of IM~Lup \citep[see Table 1 in][and \Tb{a_line_frag}]{Rosotti2023}. The latter, together with HL~Tau, display slightly higher levels of turbulence in our analysis. We discuss these two sources individually below. 
%An evident outcome of our calculations is that, for the majority of the disks, the inferred turbulence level is on the order of $10^{-4}$ in the optically thin region. Furthermore, the observed low turbulence level aligns with constraints derived from alternative methods, as discussed in \citet{Rosotti2023}. Two sources, IM~Lup and HL~Tau, display slightly higher levels of turbulence. We discuss these two sources individually below. The trend that IM~Lup has higher turbulence is consistent with constraints obtained from the direct detection of turbulence through sub-mm lines broadening \citep[e.g.,][]{FlahertyEtal2020}. Out of the six disks for which lines broadening data is available, our study includes four of them, see \Tb{a_line_frag}. 
%Notably, turbulence is not clearly detected in HD~163296, TW~Hya, and MWC~480, with the exception of IM~Lup \citep[see Table 1 in][and \Tb{a_line_frag}]{Rosotti2023}.

The low value of $\alpha_{\rm frag}$ obtained in our calculations represents the turbulent velocity among the maximum-grain-size pebbles at the midplane. Since we calculated the reference Stokes number based on a gas-to-dust mass ratio of 100, the inferred turbulence should scale linearly with the actual radial profile of the gas-to-dust mass ratio, denoted as $f_{\rm g2d}(r)$. In our default setup where we assume $f_{\rm g2d}(r)=100$, we observe a trend indicating higher turbulence at the ring location. However, we anticipate that the value of $f_{\rm g2d}(r)$ will be lower at the ring location due to dust trapping. Whether a decrease of $f_{\rm g2d}(r)$ can fully account for the increase in $\alpha_{\rm frag}$ is not definitive. In any case, the pebbles are expected to be more concentrated than the gas, hence having a gas-to-dust ratio lower than 100. This would put the true turbulence level lower than this default setup.

\begin{figure*}[thp]
    \centering
    \includegraphics[width=0.99\textwidth]{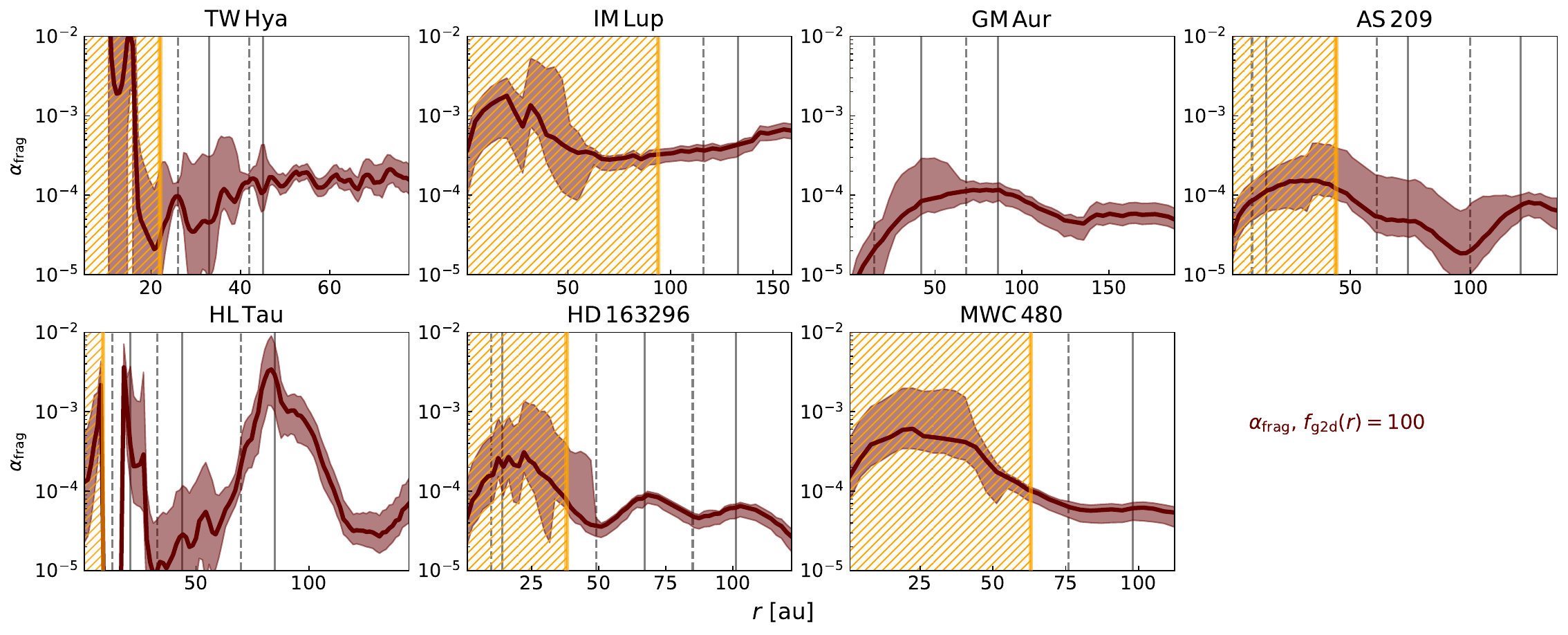}
    \caption{Calculated turbulence $\alpha_{\rm frag}$ responsible for the fragmentation velocity $v_{\rm frag} = 1$\,m\,s$^{-1}$ and gas-to-dust ratio of 100. An overall low value of $10^{-4}$ is found in five out of seven disks. {IM~Lup appears a systemically higher value of ${\sim}10^{-3}$. An increasing $\alpha_{\rm frag}$ is found in HL~Tau, which could be associated with the late infalling materials, see discussion. The infalling accretion shock impact on the disk traced by SO and SO$_2$ located at 50 to 100 au \citep{GarufiEtal2022a}, roughly consistent with the B85 ring, where is also the peak of the $\alpha_{\rm frag}$. In the the region dashed by orange lines, the data in all wavelengths used in the multi-wavelength analysis are optically thick, and therefore the results are less robust.}}
    \label{fig:alpha_r}
\end{figure*}

In the following, we provide more detailed discussions of three specific sources:
\paragraph{IM~Lup:}
Among the seven disks analyzed in this study, IM~Lup exhibits the highest turbulence level at approximately $\sim10^{-3}$, higher than any other source by one order of magnitude. This higher turbulence is consistent with constraints obtained from the direct detection of turbulence through sub-mm lines broadening \citep[e.g.,][]{FlahertyEtal2020}. This could be due to the fact that IM~Lup has been suggested to be gravitationally unstable based on the presence of spiral substructures \citep{HuangEtal2018c} and high disk masses \citep{LodatoEtal2023}.

Our measured value is systematically smaller than the gas turbulence value of $\alpha\sim0.01$ reported by \citet{FlahertyInPrep}. If true, two potential explanations could apply to this difference. First, this could indicate a significant Schmidt number (ratio of gas diffusivity to particle diffusivity) larger than the usually used value around unity \citep[e.g.,][]{YoudinLithwick2007}. Recent studies by \citet{BaehrZhu2021} propose that gravitoturbulent disks exhibit high Schmidt numbers ranging from $10-100$, which aligns with our findings. On the other hand, our measurement applies to the turbulence level at the pebble-settled midplane, yet, the line-broadening measurement is probing the line-emission surface. This geometric difference could suggest a vertical diffusivity gradient, in which the midplane is less turbulent than the upper layer, as those triggered by vertical shearing instability (VSI) \citep[e.g.,][]{StollKley2016,FlockEtal2017b,PfeilKlahr2021} {and/or magnetorotational instability (MRI) \citep[e.g.,][]{RiolsLesur2018,SimonEtal2018}.}

\paragraph{HL~Tau:}
As the youngest disk in our sample, HL~Tau is widely believed to be a Class~I disk \citep{ALMAPartnershipEtal2015}. Moreover, recent works identify clear evidence of late infalling materials onto the disk \citep{YenEtal2019,GarufiEtal2022a}. Both scenarios may suggest that grains in the outer disk have not yet reached the growth barrier, which is consistent with the overall low $a_{\rm max}$ and its decreasing trend with radius in HL~Tau (Guerra-Alvarado~et~al.~in~prep.). Therefore, our assumption that pebbles all reach the growth barrier may not apply in HL~Tau. If true, the higher turbulence value we infer in the outer region of HL~Tau should be overestimated.

{Alternatively, protostellar collapse calculations with dust coagulation show that the fragmentation threshold can be reached almost at the same time as the disk is formed \citep{TsukamotoEtal2021,KawasakiEtal2022,LebreuillyEtal2023}. If so, our method may still be valid, and thus the decreasing pebble size may indeed reflect an increasing turbulence with radii. In fact, numerical simulations of infalling materials onto disks indeed found a radially increasing turbulence \citep{LesurEtal2015,HennebelleEtal2016,HennebelleEtal2017,KuznetsovaEtal2022}. More intriguingly, the increasing $\alpha_{\rm frag}$ values we measured here are roughly consistent with the values from the simulation \citep[e.g., see Fig. 6 of ][]{KuznetsovaEtal2022}.}

\paragraph{HD~163296:}
The turbulence levels of HD163296 have been previously investigated by \citet{DoiKataoka2021} and \citet{LiuEtal2022y}. By considering the changing optical depth along the line of sight in geometrically thick but optically thin disks, both works analyzed the azimuthal brightness profiles at the two continuum rings, B67 and B100, located at 67 au and 100 au in HD163296. It was concluded that the turbulence level is low at the outer B100 ring, consistent with our findings. However, based on the observation that the brightness of the ring at the major axis is higher than the minor axis, it has been suggested that the pebble scale height is comparable to the gas scale height at the B67 ring. This implies that dust is well coupled with gas and constrains $\alpha/{\rm St}\sim1$, with similar results reported in \citet{PizzatiEtal2023}. It is important to point out that these three studies constrain only the $\alpha/{\rm St}$ ratio in the vertical direction, which depends on the Stokes number and can lead to degenerate results in terms of turbulence level estimation.

To provide a more comparable analysis, we calculate the ${\alpha_{\rm frag}}/{\rm St}$ value based on \eq{alpha_frag} for all of our sources. 
\begin{equation}\label{eq:alpha_frag_St}
    \frac{\alpha_{\rm frag}}{\rm St} 
    = \frac{4}{3\pi^2} \frac{f_{\rm g2d}^2\Sigma_{\rm d}^2}{\rho_\bullet^2 a_{\rm p}^2}\frac{v_{\rm frag}^2}{c_{\rm s}^2}
\end{equation}
However, with our default parameter values, we find that ${\alpha_{\rm frag}}/{\rm St} = 0.003$ at the 67 au ring, which is significantly smaller than unity. Therefore, we would require either a higher turbulence level or a lower Stokes number at the B67 ring to match the constraints from dust settling.

\begin{figure*}
    \centering
    \includegraphics[width=0.99\textwidth]{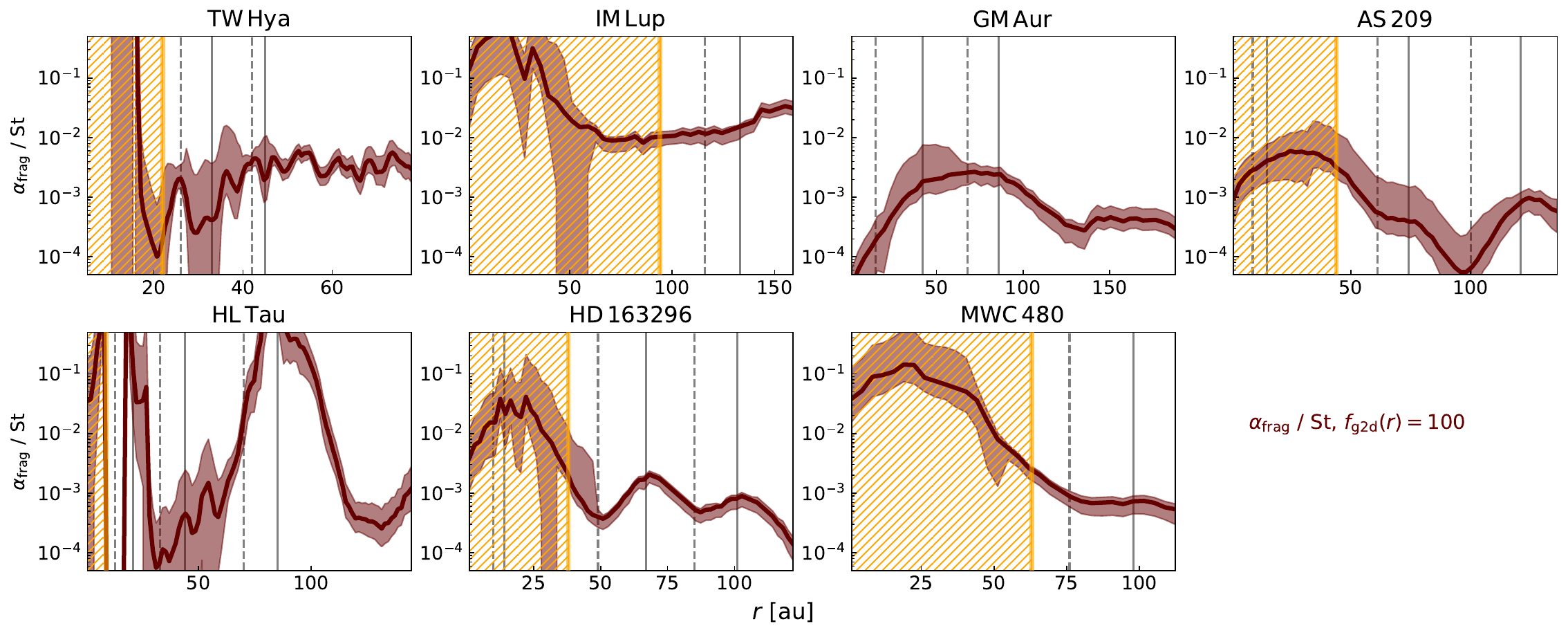}
    \caption{Calculated trapping efficiency $\alpha_{\rm frag}/\rm St$ responsible for the fragmentation velocity $v_{\rm frag} = 1$\,m\,s$^{-1}$ and gas-to-dust ratio of 100. Notations are the same as \fg{alpha_r}.}
    \label{fig:alpha_St_r}
\end{figure*}

The analysis of HD~163296's multi-wavelength continuum has been independently conducted by both \citet{SierraEtal2021} and \citet{GuidiEtal2022}. In both of these studies, the presence of a smooth radial profile in $a_{\rm max}$ is emphasized, which served as motivation for the investigation presented in our work. The smooth pebble size variation and enhanced density at the B67 ring leads to a Stokes number of $0.04$, which is lower than in other regions of the disk (see \fg{St_r}), indicating stronger aerodynamic coupling with the gas. However, this value is still too high for the $\alpha_{\rm frag}=10^{-4}$ value at the B67 ring to achieve $\alpha/{\rm St}\sim 1$, as found in other works. 

A higher gas-to-dust ratio would decrease St and increase the inferred $\alpha_{\rm frag}$, but an increase of a factor $\sim18$ would be needed to reach $\alpha/{\rm St}\sim 1$. Such an increase in the gas-to-dust ratio seems extremely unlikely given that ring substructures should be trapping the pebbles.
Another possibility to reconcile the differences in $\alpha/{\rm St}$ is that the fragmentation velocity at the B67 ring is higher.
%Therefore, to reconcile the differences between our results and the constraints from settling while keeping the low turbulence level at B100 ring, based on \eq{alpha_frag_St}, two potential solutions emerge --- larger fragmentation velocities or lower pebble internal densities at the ring location. The first can release the fragmentation barrier, and therefore relax a higher turbulence level at the ring, The second will make the Stokes number of pebbles lower at the ring location. 
Notably, the B67 ring location coincides with the CO snowline in HD~163296 \citep{ZhangEtal2021k}. Various mechanisms have been proposed to alter pebble properties at snowlines \citep{ZhangEtal2015,OkuzumiEtal2016}. %Specifically, as observed around water snowlines, water sticking efficiency is enhanced \citep{GundlachEtal2011,AumatellWurm2014}. Therefore, it is possible that similar changes in surface energy could occur near the CO snowline, leading to an increase in $v_{\rm frag}$. 
However, in order to keep a flat $a_{\rm max}$ profile at the ring, this increase in $v_{\rm frag}$ would need to spatially coincide with a simultaneous increase in $\alpha_{\rm frag}$, which seems like an unlikely coincidence.
Instead,
porosity evolution could reduce the internal pebble densities \citep{OkuzumiEtal2012}. If such evolution occurs around the CO snowline, it could also delay pebble settling \citep{OrmelEtal2007}. 
Alternatively, the inconsistency might again indicate the different turbulence levels at the midplane and puffed-up upper layers, as we suspect for IM~Lup. A final possibility is that the B67 ring is in fact composed of multiple unresolved rings, or that it has various asymmetries that make the settling analysis not valid. In fact, this ring presents an arc feature \citep{IsellaEtal2018}, which might suggest that this ring hosts a more complex structure than previously thought.

\section{Discussion}\label{sec:discussion}

\subsection{Spatial resolution}
The spatial resolution of the observations could have an effect on the derived pebble sizes and dust surface densities used in our analysis. For instance, the contrast between rings and gaps is expected to be higher if current observations do not fully resolve the rings. This could specifically happen to MWC~480, where only Band~3 and Band~6 ALMA continuum data are available, with a relatively large beam size of 35~au at Band~3. Accounting for the caveat, the trend of smooth $a_{\rm max}$ is still clear, with no significant change across the ring or gap. We obtain a low turbulence level of $10^{-4}$ in the outer disk where the disk is optically thin in both Band~3 and Band~6. 
Among the seven disks, thanks to its close distance \citep[$d = 60$~pc,][]{GaiaCollaborationEtal2021}, TW~Hya holds the highest spatial resolution, which allows the multiple wavelength observations to resolve fruitful substructure in the radial profiles of both pebble surface density and maximum grain size. A tentative enhancement of pebble size appears in the rings around 35~au and 45~au, yet the uncertainty is high. Checking the result of disk turbulence, the inferred dust diffusivity still appears very flat around the value of $10^{-4}$ when assuming $f_{\rm g2d}(r) = 100$.

Similarly, the currently limited estimates of the gas surface density profiles force us to estimate it based on the dust surface density profile and a constant gas-to-dust ratio. However, the gas-to-dust ratio is expected to vary throughout the disk and within its substructures. Robust and high spatial resolution measurements of the gas surface density distribution would allow us to circumvent these assumptions.

\subsection{Porosity of pebbles}
One limitation of this study is the uncertainty surrounding the specific properties of dust grains, particularly their composition and porosity. These factors can lead to substantial variations in terms of the internal density, opacity, and albedo of the dust \citep[e.g.,][]{MinEtal2016}.

For example, porous particles with lower internal density would result in a lower Stokes number for a given particle size. Higher porosities could thus in principle result in higher values of $\alpha_{\rm frag}$ and even higher $\alpha_{\rm frag} / {\rm St}$.

While assuming a different porosity can alter the derived profiles of $a_{\rm max}$ and dust surface density, it's worth noting that a study by \citet{ZhangEtal2023} demonstrated that the dust surface density could increase by a factor of 6 with 90\% porous particles, while $a_{\rm max}$ increased by one order of magnitude. This suggests that the impact of porosity on our results would be mitigated by a "compensation" effect between increased $a_{\rm max}$ and decreased internal density. Given the ongoing debate regarding the detailed simulation of porous pebbles, we refrain from delving deeply into the influence of pebble porosity in this work and focus on compact spherical pebbles instead.

%However, \citet{KataokaEtal2055} finds the opacities of the pebbles are strongly influenced by the with $f_{\rm fill}$, which might naturally explain the smooth radial profile of the $a_{\rm max}$ as its opacity radial profile is expected to be smooth over different size as well. 

\subsection{Bouncing barrier}
In addition to the fragmentation barrier, another dust growth barrier that can result in constant particle size is the bouncing barrier \citep{ZsomEtal2010}. As a direct consequence of bouncing, pebbles tend to attain similar sizes that remain insensitive to the viscosity and density of the gas. However, the specific size at which bouncing becomes dominant in collisions depends on sticking properties and collision velocities, which remain subjects of ongoing research in laboratory experiments.

\citet{MusiolikEtal2016a} suggested that bouncing could occur when collisional velocities fall within the range of $0.04$ to $1$\,m\,s$^{-1}$. Nevertheless, it's essential to note that the presence of different ices coating the pebbles could also influence the bouncing barrier, potentially causing it to shift \citep{Musiolik2021}. For simplicity, we omit the bouncing barrier and rely solely on the parameter $v_{\rm frag}$ to control collisional growth, aside from drifting. 

\subsection{Fragmentation velocity}
Finally, the most important parameter in this study is the fragmentation threshold defined by the fragmentation velocity $v_{\rm frag}$. As \eq{St_frag} shows, the inferred turbulence $\alpha_{\rm frag} \propto v_{\rm frag}^2$ strongly depends on the fragmentation velocity. Our primary choice of $v_{\rm frag} \leq 1$\,m\,s$^{-1}$ is widely suggested by both simulation \citep[e.g,][]{GuettlerEtal2010} and found in laboratory experiments for silicates \citep{BeitzEtal2011}, cold CO$_2$, H$_2$O and their mixtures \citet{MusiolikEtal2016a,MusiolikEtal2016b,FritscherTeiser2021}, and for organic materials \citep{BischoffEtal2020}. A lower $v_{\rm frag}$ would require significantly lower relative velocity between colliding pebbles to allow the size of the pebbles to grow to the observed size, which would push the turbulence level even lower. 
%In addition, for pebbles settled towards the midplane with no additional turbulence, the pebble layer should be sustained by Kelvin-Helmholtz instability (KHI) \citep{Sekiya1998,YoudinShu2002}. By assuming diffusion and settling are balanced in the pebble scale set by KHI and taking the Richardson number 0.1 \citep{Chiang2008}, \citet[][see their Eq.32]{JiangOrmel2021} obtain an effective turbulence coefficiency
%\begin{equation}
%    \delta_{\rm c} \simeq 1\times10^{-5}\left(\frac{\rm St}{10^{-2}}\right)\left(\frac{h_{\rm g}}{0.1}\right)^2
%\end{equation}
%\citep[cf.][]{YoudinChiang2004,TakeuchiEtal2012}, which is roughly the lower limit for the turbulence level in a protoplanetary disk.

We note that a higher $v_{\rm frag}$ (like the more commonly used value of $10\rm\,m\,s^{-1}$) would require alphas of the order of $10^{-2}$ in order to prevent the significant grain growth at the ring, as shown in \fg{a4f10_2r_A10}. 
Such high turbulence is inconsistent with current constraints in the literature {\citep[see Fig.3 of][for a recent summary]{Rosotti2023}}. 
In \App{add_model}, we explore further this scenario with additional \texttt{DustPy} simulations. In the simulation with higher $v_{\rm frag}=10$\,m\,s$^{-1}$ and higher $\alpha=10^{-2}$, even though similar maximum grain sizes appear, the radial concentration of pebbles is much weaker. 
As such, narrower/stronger gas rings would be required to form the narrow dust rings as observed by ALMA, which should lead to a strong one-to-one spatial correlation between the dust rings and gas rings. 
Yet, such a correlation does not emerge in the MAPS sources \citep{LawEtal2021a,JiangEtal2022}. 
We cannot fully exclude moderate $v_{\rm frag}=3$\,m\,s$^{-1}$ and $\alpha=10^{-3}$, though further support from the laboratory side would be required. If $v_{\rm frag}=3$\,m\,s$^{-1}$ was used, our results in \fg{alpha_r} would scale up by a factor of 9. 

\subsection{Implications for dust evolution}
The most direct implication of this work is, again, that the maximum pebble size $a_{\rm max}$ in disks (at least inside the rings) is limited by fragmentation. 
Our finding also serves as a direct support to the "constant" Stokes over the disk, as proposed for disks under coagulation/fragmentation equilibrium by \citet{BirnstielEtal2011}. 
%In that sense, this method can ease the dust evolution modeling for cases where the low mass tail of dust is less important, such as pebble accretion \citep{Ormel2017}, where the planet growth is mostly contributed by the well-settled top-heavy pebbles. 

Under this scenario, by assuming $v_{\rm frag}$ and $\alpha_{\rm frag}$, one can directly get the Stokes number of the largest pebbles, without the need to solve the complex coagulation, fragmentation module, such as in \texttt{Dustpy}. In that sense, this method can ease the dust evolution modeling for cases where the low mass tail of dust is less important, such as pebble accretion \citep{Ormel2017,JiangOrmel2023}, where the planet growth is mostly contributed by the well-settled top-heavy pebbles. The obtained grain sizes and Stokes numbers can also serve as guidance for the parameter choice in hydrodynamic simulations for experiments related to, for example, the streaming instability \citep{LiYoudin2021}, where the pebble size might have a great impact, but detailed coagulation/fragmentation would be too expensive to couple with the hydrosimulation. 

\section{Conclusion}\label{sec:conclusion}
We compare dust size properties inferred from the multi-wavelength continuum analysis for 7 protoplanetary disks, HL~Tau (Guerra-Alvarado~et~al.~in~prep.), TW~Hya \citep{MaciasEtal2021} and the 5 MAPS disks \citep{SierraEtal2021}. Different from the prediction of significant grain growth at the pressure maxima, a smooth maximum pebble size radial profile is found to be a common feature in the outer region in those samples, even at pebble ring locations. The lack of significant grain growth can be attributed to the low fragmentation barrier prevalent in the outer regions of these disks, a result in line with the recent laboratory experiments, where the catastrophic collisional fragmentation is found to happen when the relative velocity between pebbles reaches $1$\,m\,s$^{-1}$ level. Despite the uncertainty, we propose that the fact that the growth of pebbles is limited by the fragmentation barrier can be used to infer the turbulence level of the protoplanetary disks based on the pebble size and surface densities inferred from multi-wavelength continuum analysis.

Our key finding is the low turbulence level of $10^{-4}(10^{-3})$ found in 5 out of the 7 disks in our sample, when taking $v_{\rm frag} = 1(3)$\,m\,s$^{-1}$. This suggests that low-turbulence midplanes might be a common feature in protoplanetary disks. 
Two clear outliers are identified. Compared with other disks, the IM~Lup is found to host a relatively higher turbulence level, consistent with measurements from line broadening. This hints at the presence of a different mechanism triggering its turbulence, which might be due to its gravitationally unstable nature. In this disk, the turbulence level we obtain at the midplane is systematically lower than the turbulence measured at the line-emitting upper layer by line-broadening, a feature consistent with turbulence generated by VSI {and/or MRI}. 
A radially increasing trend of turbulence level is found in HL~Tau, which might indicate that pebbles have not reached the growth barrier in the outer region of the disk {or the turbulence is enhanced at $\sim$100~au, likely because of the infalling materials.}%, a limitation of our method. 

In summary, our study on pebble growth in protoplanetary disks highlights their overall fragility and their residence within the low-turbulence midplanes of protoplanetary disks, regardless of radial drifting. This contrasts with the traditional view of pebble growth being limited by radial drifting in relatively turbulent disks, where the fragmentation barrier is only reached at drifting trapping locations, such as pressure bumps. Our work underscores the significance of pebble size measurements in protoplanetary disks and emphasizes the profound impact of collisional experiments on our understanding of protoplanetary disk evolution and planet formation. Lastly, we demonstrate that multi-wavelength continuum analysis of protoplanetary disks is a unique approach for constraining not only pebble properties but also disk turbulence.

\begin{acknowledgements}
    {We thank the anonymous referee for their thoughtful and constructive comments.} 
    H.J. and E.M. are grateful to Anibal Sierra and Ke Zhang for sharing their MAPS results.
    The authors would thank Chris Ormel for useful comments. C.C.-G. acknowledges support from UNAM DGAPA-PAPIIT grant IG101321 and from CONACyT Ciencia de Frontera project ID 86372.
    This work has used \texttt{Matplotlib} \citep{Hunter2007}, \texttt{Numpy} \citep{HarrisEtal2020}{, \texttt{DustPy} \citep{StammlerBirnstiel2022}} software packages. 
\end{acknowledgements}

%\newpage
\bibliographystyle{aa} % style aa.bst
\bibliography{ads}
\begin{appendix} %First appendix
\section{Test simulations}\label{app:add_model}
We run a group of \texttt{Dustpy} simulations to test and illustrate the robustness of our major assumption --  the growth of the pebbles is under fragmentation limit in the sample disks we choose. The parameters we test are listed in \Tb{run_parameters}.

In the first group of simulations in \fg{dp_A}, we vary the gap depth amplitude $A$ in \Eq{F_gap} (\texttt{a4f01A05}, \texttt{a4f01A10}, \texttt{a4f01A20}). At the density minimum, and therefore pressure minimum, pebbles are quickly drifting away, which leads to both low densities and small pebble sizes. This becomes particularly obvious in the simulation \texttt{a4f01A20} with $A=2$. However, on the other hand, due to the low continuum flux at the gap regions, the pebble size measurements are also the most unreliable. High-spatial-resolution and high-sensitivity observation are therefore required to fully resolve the rings and gaps. Nevertheless, as the growth of pebbles is always limited by fragmentation at the ring locations, our conclusion at the rings should not be influenced by the choice of the gap amplitude.

We then test two groups of simulations with different $\alpha$ and $v_{\rm frag}$, but keeping $v_{\rm frag}^2/\alpha$ the same to maintain the fragmentation barrier almost constant. In the first group (\texttt{a4f01A00}, \texttt{a3f03A00}, \texttt{a2f10A00}), we conduct our simulation in a smooth gas disk with $A=0$, see \fg{dp_smooth}. In all of the three simulations, the pebble growth is limited by the fragmentation barrier and, since the maximum pebble sizes scale with $v_{\rm frag}^2/\alpha$, the pebbles reach the same size. This would make the initial input indistinguishable if one relies solely on the pebble properties. However, in the second group of tests (\texttt{a4f01A10}, \texttt{a3f03A10}, \texttt{a2f10A10}), we run the same $\alpha$ and $v_{\rm frag}$ combinations but in gapped disks with $A=1$. Clear differences emerge in \fg{dp_pb}. The dust concentration can only happen at the simulations with lower $\alpha\lesssim10^{-3}$. In the highly turbulent ($\alpha=10^{-2}$) simulation, although sticky pebbles ($v_{\rm frag} = 10$\,m\,s$^{-1}$) can still grow to the (sub)mm sizes observed by ALMA, dust concentration is very hard to happen, and thus no ring structures are present in the disks. 

\begin{table}[thp]
\caption{Model parameters}
\label{tab:run_parameters}
\centering
\small
\begin{tabular}{l|ccc}
\hline\hline
run-id & alpha & fragmentation velocity & gap depth amplitude \\
    &   & $\rm m\,s^{-1}$  &   \\
\hline
\texttt{a4f01A00}    &   $10^{-4}$   & 1  &  0 \\
\texttt{a4f01A10}    &   $10^{-4}$   & 1  &  1 \\
\texttt{a4f03A10}    &   $10^{-4}$   & 3  &  1 \\
\texttt{a4f10A10}    &   $10^{-4}$   & 10 &  1 \\
\texttt{a4f01A05}    &   $10^{-4}$   & 1  & 0.5\\
\texttt{a4f01A20}    &   $10^{-4}$   & 1  &  2 \\
\texttt{a3f03A00}    &   $10^{-3}$   & 3  &  0 \\
\texttt{a3f03A10}    &   $10^{-3}$   & 3  &  1 \\
\texttt{a2f10A00}    &   $10^{-2}$   & 10 &  0 \\
\texttt{a2f10A10}    &   $10^{-2}$   & 10 &  1 \\
\hline
\end{tabular}
\end{table}

\begin{figure*}[tbp]
\centering  
\subfiguretopcaptrue
\vspace{-0.3 cm}
\subfigure{
\includegraphics[height=0.383\textwidth]{fig/sim/a_Sigma.pdf}}\hspace{-8pt}
\setcounter{subfigure}{0}
\hfill
\subfigure[$A = 0.5$]{\label{fig:a4f1_2r_A05}
\includegraphics[height=0.383\textwidth]{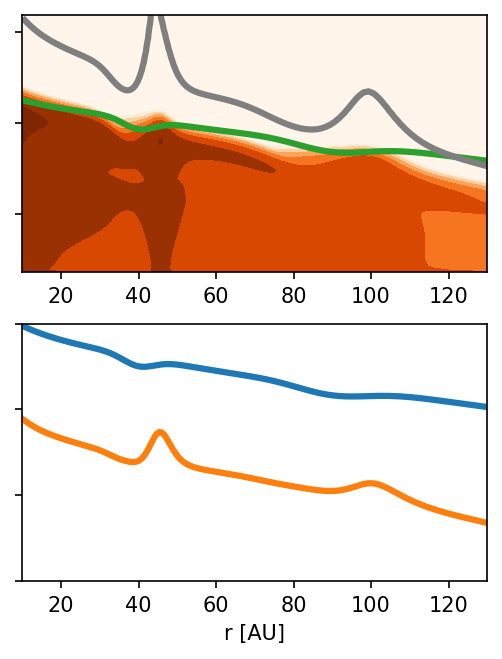}}\hfill\hspace{-10pt}
\subfigure[$A = 1$]{\label{fig:a4f1_2r_A10_2}
\includegraphics[height=0.383\textwidth]{fig/sim/a4f1_2r_A10.png}}\hfill\hspace{-10pt}
\subfigure[$A = 2$]{\label{fig:a4f1_2r_A20}
\includegraphics[height=0.383\textwidth]{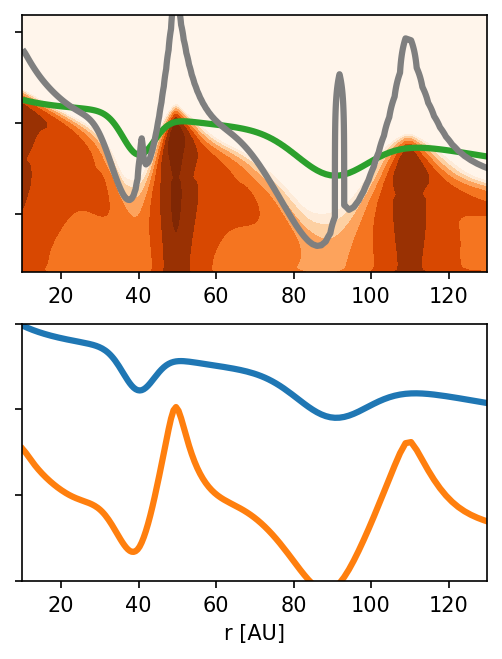}}\hfill\hspace{-10pt}
\subfigure{
\includegraphics[height=0.383\textwidth]{fig/sim/delta_d.pdf}}\hfill
\hfill
\subfiguretopcapfalse
\vspace{-.5 cm}
\caption{Simulation outputs at $t=1$~Myr for runs with different gap amplitude $A$. The fragmentation velocities $v_{\rm frag} = 1\rm\,m\,s^{-1}$ and the dust diffusivities are set to be $10^{-4}$ for all three runs. Labels are the same as in \fg{dp_v_frag}.}
\label{fig:dp_A}
\end{figure*}

\begin{figure*}[tbp]
\centering  
\subfiguretopcaptrue
\vspace{-0.3 cm}
\subfigure{
\includegraphics[height=0.383\textwidth]{fig/sim/a_Sigma.pdf}}\hspace{-8pt}
\setcounter{subfigure}{0}
\hfill
\subfigure[$\alpha = 10^{-4}$, $v_{\rm frag} = 1$\,m\,s$^{-1}$]{\label{fig:a4f1}
\includegraphics[height=0.383\textwidth]{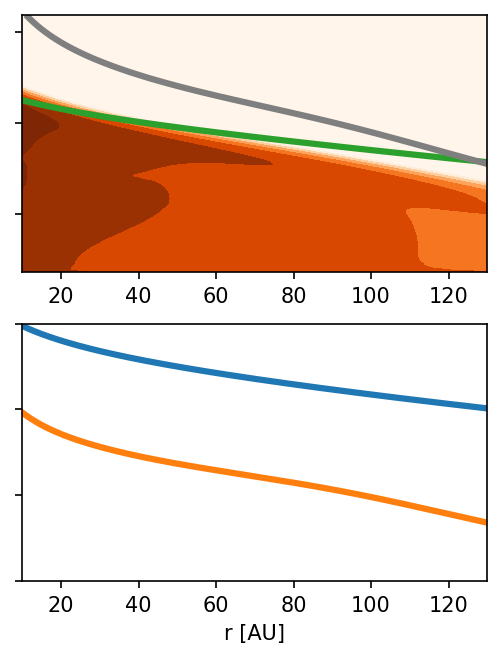}}\hfill\hspace{-10pt}
\subfigure[$\alpha = 10^{-3}$, $v_{\rm frag} = 3$\,m\,s$^{-1}$]{\label{fig:a3f3}
\includegraphics[height=0.383\textwidth]{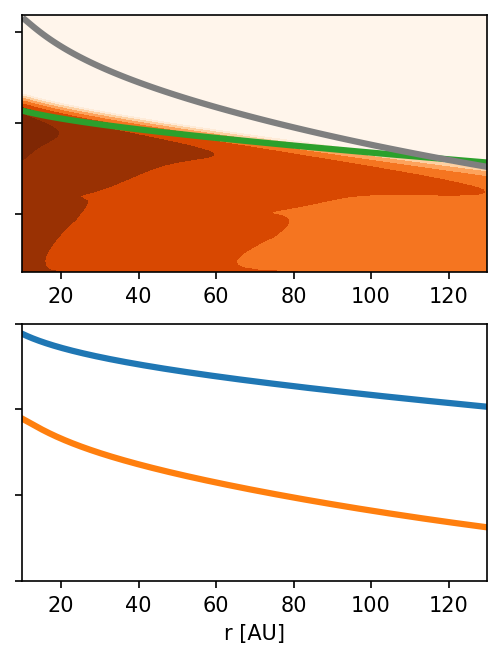}}\hfill\hspace{-10pt}
\subfigure[$\alpha = 10^{-2}$, $v_{\rm frag} = 10$\,m\,s$^{-1}$]{\label{fig:a2f10}
\includegraphics[height=0.383\textwidth]{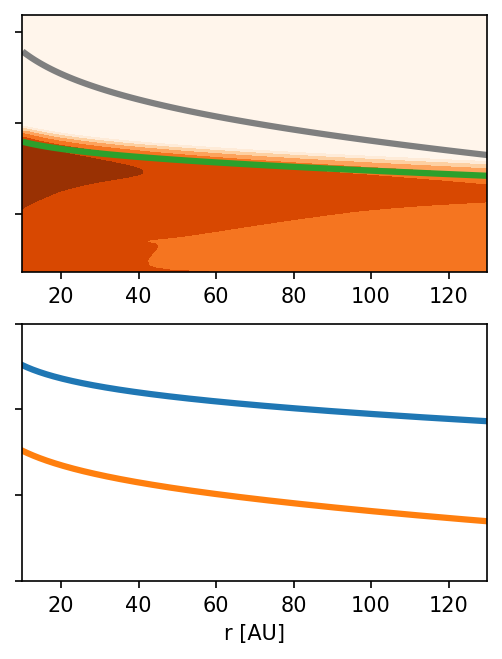}}\hfill\hspace{-10pt}
\subfigure{
\includegraphics[height=0.383\textwidth]{fig/sim/delta_d.pdf}}\hfill
\hfill
\subfiguretopcapfalse
\vspace{-.5 cm}
\caption{Simulation outputs at $t=1$~Myr for runs with different dust diffusivity and fragmentation velocity in smooth disks ($A=0$). The outputs are indistinguishable. Labels are the same as in \fg{dp_v_frag}.}
\label{fig:dp_smooth}
\end{figure*}

\begin{figure*}[tbp]
\centering  
\subfiguretopcaptrue
\vspace{-0.3 cm}
\subfigure{
\includegraphics[height=0.383\textwidth]{fig/sim/a_Sigma.pdf}}\hspace{-8pt}
\setcounter{subfigure}{0}
\hfill
\subfigure[$\alpha = 10^{-4}$, $v_{\rm frag} = 1$\,m\,s$^{-1}$]{\label{fig:a4f1_2r_A10_3}
\includegraphics[height=0.383\textwidth]{fig/sim/a4f1_2r_A10.png}}\hfill\hspace{-10pt}
\subfigure[$\alpha = 10^{-3}$, $v_{\rm frag} = 3$\,m\,s$^{-1}$]{\label{fig:a3f3_2r_A10}
\includegraphics[height=0.383\textwidth]{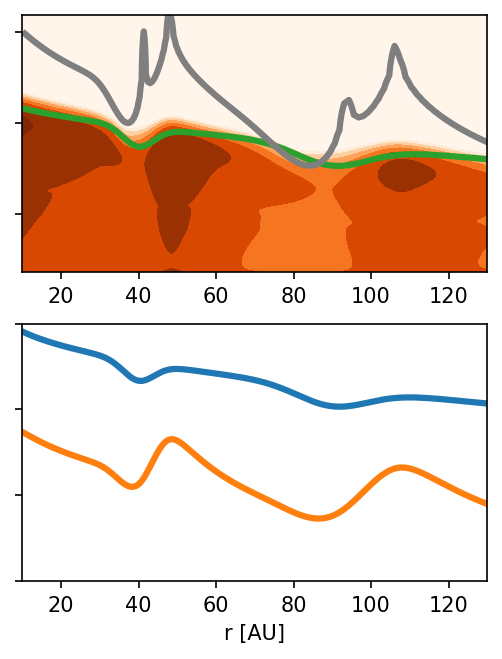}}\hfill\hspace{-10pt}
\subfigure[$\alpha = 10^{-2}$, $v_{\rm frag} = 10$\,m\,s$^{-1}$]{\label{fig:a2f10_2r_A10}
\includegraphics[height=0.383\textwidth]{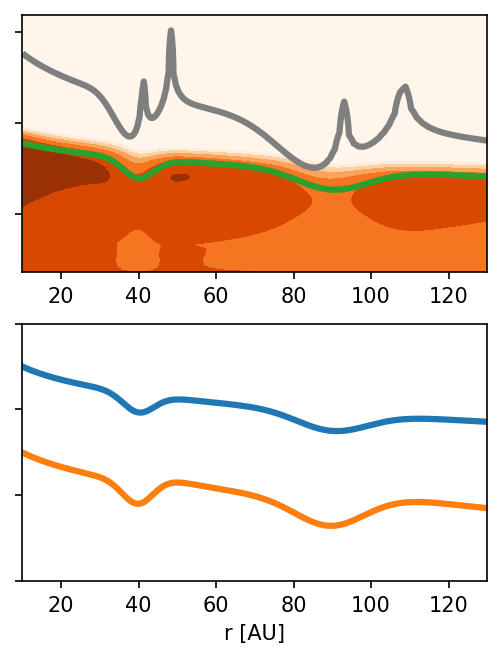}}\hfill\hspace{-10pt}
\subfigure{
\includegraphics[height=0.383\textwidth]{fig/sim/delta_d.pdf}}\hfill
\hfill
\subfiguretopcapfalse
\vspace{-.5 cm}
\caption{Simulation outputs at $t=1$~Myr for runs with different dust diffusivity and fragmentation velocity in gaped disks ($A=1$). Other setup parameters are the same as \fg{dp_smooth}}
\label{fig:dp_pb}
\end{figure*}

\iffalse
\section{Inferred turbulence with $v_{\rm frag} = 3\rm\,m\,s^{-1}$}\label{app:alpha_3}

\begin{figure*}
    \centering
    \includegraphics[width=0.99\textwidth]{fig/obs/alpha_r_3.pdf}
    \caption{Calculated turbulence $\alpha_{\rm frag}$ responsible for the fragmentation velocity $v_{\rm frag} = 3$\,m\,s$^{-1}$ and gas-to-dust ratio of 100.}
    \label{fig:alpha_r_3}
\end{figure*}

\begin{figure*}
    \centering
    \includegraphics[width=0.99\textwidth]{fig/obs/alpha_St_r_3.pdf}
    \caption{Calculated trapping efficiency $\alpha_{\rm frag}/\rm St$ responsible for the fragmentation velocity $v_{\rm frag} = 3$\,m\,s$^{-1}$ and gas-to-dust ratio of 100.}
    \label{fig:alpha_St_r_3}
\end{figure*}
\fi

\end{appendix}
\end{document}